\documentstyle[12pt]{report}
\begin{document}

\newcounter{popnr}\setcounter{popnr}{\value{equation}}
\addtocounter{popnr}{1}
\newcommand{\alpheqn}{\setcounter{equation}{0}\renewcommand{\theequation}{\arabic{chapter}.\arabic{popnr}\alph{equation}}}
\newcommand{\reseteqn}{\setcounter{equation}{\value{popnr}}\addtocounter{popnr}{
1}\renewcommand{\theequation}{\arabic{chapter}.\arabic{equation}}}
\renewcommand{\thesection}{\arabic{chapter}.}

\newtheorem{teorem}{Theorem}[chapter]
\newtheorem{proposisjon}[teorem]{Proposition}
\newtheorem{definisjon}[teorem]{Definition}
\newtheorem{lemma}[teorem]{Lemma}
\newtheorem{korollar}[teorem]{Corollary}
\newtheorem{setning}[teorem]{Setning}
\newenvironment{Proof}{{\em Proof:} \par}%
{\hfill\framebox[1em]{\rule[-.3ex]{0em}{1.0ex}\vspace{3ex}}\par}

\newenvironment{eksempel}{\par \addtocounter{teorem}{1}%
{\bf Eksempel\ \theteorem} \par}%
{\hfill\framebox[1em]{\rule[-.3ex]{0em}{1.0ex}\vspace{3ex}}\par}

\makeatletter
\newcommand{\verbinput}[1]{{\@verbatim\frenchspacing\@vobeyspaces
                 \input{#1}\endtrivlist}}
\def\abstract{\if@twocolumn
\section*{Abstract}
\else \small
\begin{center}
{\bf Abstract\vspace{-.5em}\vspace{0pt}}
\end{center}
\quotation
\fi}
\def\endabstract{\if@twocolumn\else\endquotation\fi}
\makeatother
\newcommand{\beq}{\begin{eqnarray}}
\newcommand{\eeq}{\end{eqnarray}}
\newcommand{\beqq}{\begin{eqnarray*}}
\newcommand{\eeqq}{\end{eqnarray*}}
\newcommand{\un}{\underline}
\newcommand{\bm}[1]{\mbox{\boldmath $#1$}}
\newcommand{\pr}{Poincar\'{e}}
\newcommand{\ukp}{U_\kappa ({\cal P}_{4;1})}
\def\po{{\kern 0.55em \raise .27ex \hbox{$\scriptscriptstyle +$}}{\kern -0.55em
{\supset}}}
\thispagestyle{empty}
\begin{center}
{\Large \bf Twisted Classical Poincar\'{e} Algebras}\\[1cm]
Jerzy Lukierski$^{1),4)}$, Henri Ruegg$^{4)}$ and Valerij N. Tolstoy$^{2),4)}$
\\[5mm]
Dept. de Physique Theorique, Universit\'{e} de Geneve,\\
24, qui Ernest-Ansermet, 1211 Geneve 4, Switzerland\\[8mm]
Anatol Nowicki$^{3)}$\\[5mm]
Physikalisches Inst., Universit\"{a}t Bonn, Nussallee 12, 53115 Bonn,
Germany\\ \end{center}
\vspace*{1cm}
\centerline{\bf Abstract}
\vspace*{0.4cm}
We consider the twisting of Hopf structure for classical
enveloping algebra $U(\hat{g})$, where $\hat{g}$ is the inhomogenous rotations
algebra, with
explicite formulae given for $D=4$ Poincar\'{e} algebra $(\hat{g}={\cal P}_4).$
The comultiplications of
twisted $U^F({\cal P}_4)$ are obtained by conjugating primitive classical
coproducts
by $F\in U(\hat{c})\otimes U(\hat{c}),$ where $\hat{c}$ denotes any Abelian
subalgebra of ${\cal P}_4$,
and the universal $R-$matrices for  $U^F({\cal P}_4)$ are triangular. As an
example we show that the
quantum deformation of Poincar\'{e} algebra recently proposed by Chaichian and
Demiczev is a
twisted classical Poincar\'{e} algebra. The interpretation of twisted
Poincar\'{e} algebra as describing
relativistic symmetries with clustered 2-particle states is proposed.
\vspace{\fill}

---------------------------\\
$^{1)}$ On leave of absence from the Institute for Theoretical Physics,
University of
Wroclaw, pl. Maxa Borna 9, 50-204 Wroclaw, Poland.\\
$^{2)}$ On leave of absence from Institute of Nuclear Physics, Moscow State
University, 119899 Moscow, Russia.\\
$^{3)}$ On leave of absence from the Institute of Physics, Pedagogical
University, Pl. S{\l}owia\'{n}ski 6, 65-029 Zielona G\'{o}ra, Poland.\\
$^{4)}$ Partially supported by Swiss National Science Foundation and by OFES:
contract No. 93.0083 \\
\newpage
\section{Introduction}
Let us consider Poincar\'{e} algebra ${\cal P}_4$ with the generators
$\hat{g}=(P_\mu,M_{\mu \nu  })$ as
a classical Hopf algebra. We supplement the well-known algebraic relations

\beqq
[M_{\mu \nu  },M_{\rho\tau }]=i(g_{\mu \tau } M_{\nu  \rho }-g_{\nu
\tau}M_{\mu \rho}+g_{\nu  \rho}M_{\mu \tau}
-g_{\mu \rho}M_{\nu  \tau})
\eeqq
\beq
[M_{\mu \nu  },P_\rho ]=i(g_{\nu  \rho}P_\mu  - g_{\mu \rho}P_\nu  )
\eeq
\beqq
[P_\mu ,P_\nu  ]=0
\eeqq
by the ``primitive '' coproduct relations\\
\parbox{12cm}{
\beqq
\Delta_0(M_{\mu \nu  })=M_{\mu \nu  }\otimes 1 + 1 \otimes M_{\mu \nu  }
\eeqq
\beqq
\Delta_0(P_\mu )=P_\mu \otimes 1 + 1\otimes P_\mu
\eeqq}\parbox{1.5cm}{\beq \eeq}\\
and the antipode $S_0(\hat{g})=-\hat{g}\quad (\hat{g}\in {\cal P}_4)$. The
relations (1.1) lead to the well known Wigner theory of representations
of Poincar\'{e} algebra [1,2] which are spanned by the Hilbert vectors
$|m,s;p_\mu ,S_3>$,
where $m$ and $s$ describe respectively the eigenvalues of mass and
relativistic
spin (Pauli-Lubanski) Casimir, $p_\mu $ is the fourmomentum and $S_3\quad
(-S\leq S_3 \leq S)$ describe the spin
projection values. The coproduct formula dictates how to calculate the action
of
the Poincar\'{e} genarators on tensor product.

The quantum deformations of Poincar\'{e} algebra are described by the
modi\-fi\-ca\-tions of the relations
(1.1-2) preserving the Hopf algebra structure (for general framework see e.g.
[3,4]).
In this paper we would like to consider the mildest quantum deformations of
(1.1-2) obtained by
the twisting procedure [5-9]. Following Drinfeld [5] two Hopf algebras ${\cal
A}=(A,\Delta,S,\varepsilon )$
and ${\cal A}^F=(A,\Delta^F,S^F,\varepsilon )$ are related by twisting
if there exists an invertible function $F=\sum_i f_i\otimes f^i\in {\cal
A}\otimes {\cal A}$\footnote{
Strictly speaking we consider below $F$ belonging to an extension of ${\cal
A}\otimes {\cal A}$.}$^)$
satisfying the ``cocycle'' condition [5,7,8]
\beq
F_{23}(1\otimes\Delta )F = F_{12}(\Delta\otimes 1)F
\eeq
and $(\varepsilon\otimes 1)F=(1\otimes\varepsilon )F=1.$ In such a case
$\Delta^F$ and
$\Delta$ are related as follows $(a\otimes b \cdot c\otimes d = ac \otimes bd)$
\setcounter{popnr}{\value{equation}}
\addtocounter{popnr}{1}
\alpheqn
\beq
\Delta^F(a)=F\cdot \Delta (a)\cdot F^{-1}
\eeq
Introducing $U=\sum_{i} f_i \cdot S(f^i)$ one obtains also that
\beq
S^F(a)=US(a)U^{-1}
\eeq
\reseteqn
If ${\cal A}$ is the quasitriangular Hopf algebra and the relations
(1.3) are replaced by [5,6]
\beq
(\Delta \otimes 1)F=F_{13}F_{23}\quad\quad\quad (1\otimes \Delta
)F=F_{13}F_{12}
\eeq
the universal $R-$matrices for ${\cal A}$ and ${\cal A}^F$ are related by the
formulae
$(\tilde{F}=\sigma\cdot F=\sum_i f^i\otimes f_i)$
\beq
R^F=F^{-1}\cdot R \cdot \tilde{F}
\eeq
For the complex simple Lie algebras $\hat{g}$ there were considered twistings
de\-scri\-bed by
\beq
F=\exp f \quad\quad\quad f\in \hat{c}\otimes \hat{c}
\eeq
where $\hat{c}$ is the commutative subalgebra of $\hat{g}$ (Cartan subalgebra
in
[6], Borel subalgebra in [8]). Indeed it is
easy to check that if $f\in \hat{c}\otimes\hat{c}$, and $\hat{c}$ is abelian,
the
conditions (1.5) are valid.

In this paper we shall consider the twisting of physically important case
of inhomogeneous rotation
algebras $\hat{g}=O(D-k,k)\po\ T_D$, in particular the D=4 \pr\ algebra
$\hat{g}=O(3,1)\po\ T_4$. In such nonsimple algebrs one can select the
commutative
subalgebra $C_m$ in several ways, e.g.\\
a) Cartan subalgebra $(h_1,\ldots ,h_n)\quad (n=\frac{D}{2}\;\; \mbox{ for D
even,} \;n=\frac{D-1}{2}
 \;\;\mbox{ for D odd})$\\
b) Translation generators $(P_1\ldots P_D)$\\
c) ``Mixed'' Cartan--translation algebra $C_k\quad k\leq\frac{N}{2})$
\beq
C_k=(h_1\ldots h_k,P_{2k+1}\ldots P_D)
\eeq
The aim of this paper is\\
a) to describe the twistings of $U_q({\cal P}_4)$ depending on Cartan
generators and
translation generators.\\
b) to provide an interesting example.

In Sect.2 we shall consider in explicite way the twisted D=4 \pr\ algebras
$U^F({\cal P}_4)$ with
the choice of the algebra $\hat{g}$ (see (1.7)) described by the formula (1.8)
with
$k=0,1,2.$ Further generalization in the presence of central generators $Z_i$
$([Z_i,\hat{g}]=0$ for $\hat{g}\in {\cal P}_4$)
is also given. In Sect.3 we shall discuss as an example of classical
twisted \pr\ algebra the
quantum \pr\ algebra considered recently by Chaichian and Demiczev [10].
In Sect.4 we shall discuss
the elements of the representation theory of twisted \pr\ algebras, and present
an outlook: some generalizations as well unsolved problems.
\addtocounter{chapter}{1}
\section{Twisting of the classical Poincar\'{e} algebra.}
\setcounter{equation}{0}
Let us denote the basis of the commutative algebra
$\hat{c} \ (F\in\hat{c}\otimes\hat{c})$ by
$(c_1\ldots c_n).$ We define
\beq
F=F_+F_-\quad\quad\quad F_\pm =\exp f_\pm
\eeq
where $F_\pm =\pm\sigma\cdot f_\pm$ ($\sigma$ is the exchange map:
$\sigma(c_i\otimes c_j)=
c_j\otimes c_i$), and
\beq
f^{(\pm )} =\frac{1}{2}\alpha^{(\pm )}_{ij} (c_i\otimes c_j \pm c_j\otimes c_i)
\eeq
i.e. one can assume that $\alpha_{\pm ij}=\pm\alpha_{\pm ji}.$

If we twist the coproducts of classical Lie algebra
we obtain from the
commutativity of $\hat{c}$ that
\beq
U=\sum f_i\otimes S(f^i)=\exp( -\alpha_{+ ij}c_i c_j)
\eeq
and after using (1.6) the $R-$matrix takes the particular form:
\beq
R=\exp (-2f_-)=(F_-)^{-2}
\eeq
The formulae for the coproduct $\Delta^F$ depend on the particular choice of
the
algebra $\hat{c}$.
We shall further specify our algebra for the case of classical \pr\ algebra
(1.1),
and we shall consider following three types of the twist function:\\
a) $\hat{c}=(M_3=M_{12},\;\; N_3=M_{30})$\\
We postulate\\
\parbox{12cm}{\beqq
f_+=\alpha_+M_3\otimes M_3+\beta_+(M_3\otimes N_3+N_3\otimes
M_3)+\gamma_+N_3\otimes N_3
\eeqq
\beqq
f_-=\beta_-(M_3\otimes N_3 - N_3 \otimes M_3)\eeqq}\parbox{1.5cm}{\beq \eeq}\\
One gets $(M_i\equiv \frac{1}{2}\epsilon_{ijk}M_{jk};\,M_\pm \equiv M_1 \pm
iM_2;\; N_i\equiv
M_{i0};\, P_\pm \equiv P_1 \pm iP_2)$\\
\parbox{12cm}{\beqq
\Delta^F (M_\pm)= M_\pm \otimes e^{\pm A_1}\cos(B_1) + e^{\pm A_2}\cos(B_2)
\otimes M_\pm
\eeqq
\beqq
\pm N_\pm \otimes e^{\pm A_1}\sin(B_1) \pm e^{\pm A_2}\sin(B_2)\otimes
N_\pm \eeqq
\beqq
\Delta^F (M_3)=M_3\otimes 1 + 1\otimes M_3
\eeqq
\beqq
\Delta^F(N_\pm)= N_\pm \otimes e^{\pm A_1}\cos(B_1) + e^{\pm A_2}\cos(B_2)
\otimes N_\pm
\eeqq
\beqq
\mp M_\pm \otimes e^{\pm A_1}\sin(B_1) \mp e^{\pm A_2}\sin
(B_2)\otimes M_\pm
\eeqq
\beqq
\Delta^F(N_3)=N_3\otimes 1 + 1\otimes M_3
\eeqq}\parbox{1.5cm}{\beq \eeq}\\
\beqq
\Delta^F(P_\pm)= P_\pm \otimes e^{\pm A_1} + e^{\pm A_2}\otimes P_\pm
\eeqq
$$
$$
\beqq
\Delta^F(P_3)= P_3 \otimes \cos(B_1) + \cos(B_2)\otimes P_3 +
iP_0 \otimes \sin(B_1) + i\sin(B_2)\otimes P_0
\eeqq
\beqq
\Delta^F(P_0)= P_0 \otimes \cos(B_1) + \cos(B_2)\otimes P_0 +
iP_3\otimes \sin(B_1) + i\sin(B_2)\otimes P_3
\eeqq
where\medskip\\
\hspace*{10em}$A_k=\alpha_{+} M_3 + (\beta_{+}
-(-1)^k\beta_{-})N_3$\medskip\\
\hspace*{10em}$B_k=\gamma_{+} N_3 + (\beta_{+} +
(-1)^k\beta_{-})M_3$\bigskip\\
b) $\hat{c}=(M_3=M_{12}, P_3, P_0)$\\
We assume that $(r,s=3,0)$\\
\parbox{12cm}{\beqq
f_+ = \alpha_+ M_3\otimes M_3 +\delta^r_+ (M_3\otimes P_r + P_r \otimes M_3)
+\rho^{rs}_+  P_r\otimes P_s
\eeqq
\beqq
f_-=\delta^r_-(M_3\otimes P_r - P_r\otimes M_3)
\eeqq}\parbox{1.5cm}{\beq \eeq}\\
One obtains\\
\parbox{12cm}{
\beqq
\Delta^F(M_\pm)=M_\pm \otimes e^{\pm A_1} + e^{\pm A_2}\otimes M_\pm
\pm P_\pm \otimes B_3e^{\pm A_1} \pm e^{\pm A_2}C_3 \otimes P_\pm
\eeqq
\beqq
\Delta^F(M_3)=M_3\otimes 1 + 1 \otimes M_3
\eeqq
\beqq
\Delta^F(N_\pm)= N_\pm \otimes e^{\pm A_1} + e^{\pm A_2} \otimes N_\pm
-i{P_\pm \otimes B_0 e^{\pm A_1}  + C_0 e^{\pm A_2}\otimes P_\pm}
\eeqq
\beqq
\Delta^F(N_3)=N_3 \otimes 1 + 1 \otimes N_3 -i{P_3 \otimes B_0 + C_0
\otimes P_3 + P_0 \otimes B_3 + C_3 \otimes P_0}
\eeqq
\beqq
\Delta^F(P_1)=P_1 \otimes \cosh(A_1)+ \cosh(A_2) \otimes P_1
+ iP_2 \otimes \sinh(A_1) + i\sinh(A_2) \otimes P_2
\eeqq
\beqq
\Delta^F(P_2)=P_2 \otimes \cosh(A_1) + \cosh(A_2) \otimes P_2
-iP_1 \otimes \sinh(A_1) -i\sinh(A_2)\otimes P_1
\eeqq}\parbox{1.5cm}{\beq \eeq}\\
\beqq
\Delta^F(P_3)=P_3 \otimes 1 + 1 \otimes P_3
\eeqq
\beqq
\Delta^F(P_0)=P_0 \otimes 1 + 1 \otimes P_0
\eeqq
where:\bigskip\\
\hspace*{4em}$A_1 = \alpha_+ M_3 + (\delta_{+}^{r} + \delta_{-}^{r})P_r
\qquad B_r=(\delta_{+}^{r} - \delta_{-}^{r})M_3 +
\rho_{+}^{rs}P_s$\bigskip\\
\hspace*{4em}$A_2= \alpha_+ M_3 + (\delta_{+}^{r} - \delta_{-}^{r})P_r
\qquad C_r=\rho_{+}^{sr}P_s + (\delta_{+}^{r} +
\delta_{-}^{r})M_3$\bigskip\\

c) $\hat{c}=(P_1, P_2, N_3=M_{30})$\\
Putting (a,b=1,2)\\
\parbox{12cm}{
\beqq
f_+=\rho^{ab}_+ P_a\otimes P_a + \xi^a_+ (N_3\otimes P_a + P_a \otimes N_3)
+\gamma_+ N_3\otimes N_3
\eeqq
\beqq
f_- = \xi^a_- (N_3\otimes P_a - P_a \otimes N_3)
\eeqq}\parbox{1.5cm}{\beq \eeq}\\
one gets\\
\parbox{12cm}{
\beqq
\Delta^F(M_\pm)=M_\pm \otimes \cos(A_1) + \cos(A_2)\otimes M_\pm
\pm \{N_\pm \otimes \sin(A_1)+\sin(A_2)\otimes N_\pm\}
\eeqq
\beqq
\mp\{P_3 \otimes (B_1\pm iB_2)\cos(A_1) + (C_1\pm iC_2)\cos(A_2)\otimes
P_3\}
\eeqq
\beqq
\mp i\{P_0\otimes (B_1\pm iB_2)\sin(A_1) + (C_1 \pm
iC_2)\sin(A_2)\otimes P_0\}
\eeqq
\beqq
\Delta^F(M_3)=M_3 \otimes 1 + 1\otimes M_3
\eeqq
\beqq
- i\{P_2
\otimes B_1 + C_1 \otimes P_2\} + i\{P_1\otimes B_2 + C_2\otimes P_1\}
\eeqq
\beqq
\Delta^F(N_\pm)= N_\pm \otimes \cos(A_1) + \cos(A_2)\otimes N_\pm
\mp \{ M_\pm \otimes \sin(A_1) + \sin(A_2) \otimes M_\pm \}
\eeqq
\beqq
-i \{ P_0 \otimes (B_1 \pm iB_2)\cos(A_1) + (C_1 \pm iC_2)\cos(A_2)
\otimes P_0 \}
\eeqq}\parbox{1.5cm}{\beq \eeq}\\
\beqq
+  P_3\otimes (B_1 \pm iB_2)\sin(A_1) + (C_1 \pm iC_2)\sin(A_2)
\otimes P_3
\eeqq
\beqq
\Delta^F(N_3)= N_3 \otimes 1 + 1 \otimes N_3
\eeqq
\beqq
\Delta^F(P_\pm)= P_\pm \otimes 1 + 1 \otimes P_\pm
\eeqq
\beqq
\Delta^F(P_3)= P_3 \otimes \cos(A_1) + \cos(A_2) \otimes P_3
+ iP_0 \otimes \sin(A_1) + i\sin(A_2) \otimes P_0
\eeqq
\beqq
\Delta^F(P_0)=P_0 \otimes \cos(A_1) + \cos(A_2) \otimes P_0
+ iP_3 \otimes \sin(A_1) + i\sin(A_2) \otimes P_3
\eeqq
where\\
\beqq
B_1 \pm iB_2 = (\rho_{+}^{1b} \pm i \rho_{+}^{2b})P_b
+(\xi_{+}^1 \pm i\xi_{+}^2 - (\xi_{-}^1 \pm i\xi_{-}^2))N_3
\eeqq
\beqq
C_1 \pm iC_2=(\rho_{+}^{1b} \pm i\rho_{+}^{2b})P_b
+ (\xi_{+}^1 \pm i\xi_{+}^2 + (\xi_{-}^1\pm i\xi_{-}^2))N_3
\eeqq

d) $\hat{c}=(P_1,P_2,P_3,P_0)$\\
one can write $(\rho_{\pm}^{\mu \nu  }=\pm \rho^{\nu  \mu }_\pm )$\\
\parbox{12cm}{
\beqq
f_+=\rho^{\mu \nu  }_+  (P_\mu \otimes P_\nu   + P_\nu  \otimes P_\mu  )
\eeqq
\beqq
f_-=\rho^{\mu \nu  }_- (P_\mu \otimes P_\nu   - P_\nu   \otimes P_\mu  )
\eeqq}\parbox{1.5cm}{\beq \eeq}\\
Because the split Casimir
\beq
C^{\mbox{{\small split}}}_2\equiv \Delta (P_\mu  P^\mu )-P_\mu  P^\mu \otimes 1
- 1\otimes P_\mu P^\mu =
2P_\mu \otimes P^\mu
\eeq
commutes with $\Delta(\hat{a})$ for any $\hat{a}\in U({\cal P}_4),$ one
can assume further that $\rho^{\mu \nu  }_+\eta_{\mu \nu  }=\rho^\mu _\mu =0$
($\eta_{\mu \nu  }=\mbox{diag} (1,-1,-1,-1))$.

The formulae for the coproduct take the form\\
\parbox{12cm}{
\beqq
\Delta^F(M_{\mu \nu  })=M_{\mu \nu  }\otimes 1 + 1\otimes M_{\mu \nu
}+(\alpha_{+\mu }\; ^\rho P_\nu  -
-\alpha_{+\nu}\;^\rho P_\mu )\otimes P_\rho
\eeqq
\beqq
+ P_\rho \otimes (\alpha_{+\mu }\;^\rho P_\nu
-\alpha_{+\nu}\;^\rho P_\mu )+
+(\alpha_{-\mu }\;^\rho P_\nu  -\alpha_{-\nu}\;^\rho P_\mu )\otimes
P_\rho -
\eeqq
\beqq
-P_\rho \otimes (\alpha_{-\mu }\;^\rho P_\nu   -\alpha_{-\nu}\;^\rho
P_\mu ) \eeqq
\beqq
\Delta^F (P_\mu )=P_\mu \otimes 1 + 1\otimes P_\mu
\eeqq}\parbox{1.5cm}{\beq \eeq}\\

In general we assume that the \pr\ algebra is the complex one, and the twist
function parameters
are also complex. The reality condition imposed on the \pr\ generators imply
the reality conditions for the coefficient in the formulae (2.5), (2.7), (2.9)
and (2.11).
For simplicity we shall consider the last example of the  twist function, given
by (2.11).
It is known that if the real structure is an antihomomorphism in the algebra
sector, still one can impose on the generators of twisted \pr\ algebra two
types
of reality
conditions [10,17]:\\
a) Standard one, denoted in [17] by +. In the case of the formulae (2.13) one
obtains
\setcounter{popnr}{\value{equation}}
\addtocounter{popnr}{1}
\alpheqn
\beq
(\Delta (M_{\mu \nu  }))^+=\Delta (M_{\mu \nu  } )\Longrightarrow
\alpha^{\rho\tau} \quad \mbox{real}
\eeq
b) nonstandard one, used e.g. in [18], and denoted in [10] by $\oplus$. In such
a case
\beq
(\Delta (M_{\mu \nu  }))^\oplus =\Delta (M_{\mu \nu  }) \Longrightarrow
\alpha^{\rho\tau}=(\alpha^{\tau\rho})^*
\eeq
\reseteqn
i.e. the matrix $\alpha\equiv (\alpha^{\rho\tau})$ is Hermitean.

Finally we consider the extension of $\hat{g}$ by an Abelian algebra $\hat{z}$
($\hat{g}\rightarrow \hat{g}\oplus\hat{z}$), with $z_A\;\;\; (A=1,\ldots ,m)$
describing
the central charges. The formulae (2.2) determining twist function can  be
extended
as follows:
\beq
f_\pm \rightarrow f^{(z)}_\pm = f_\pm +\frac{1}{2}\beta_{\pm iA} (c_i\otimes
Z_A
\pm Z_A \otimes c_i)
\eeq
The candidates for $Z_A$ are the  central charges as well as the Casimir
operators.
As an example we shall consider the case d) with one central charge $Z$, i.e.
we
assume
that the formulae (2.11) is extended as follows:
\beq
f_\pm \rightarrow f_\pm^{(z)} = f_\pm + \rho^\mu _\pm (P_\mu  \otimes Z \pm
Z\otimes P_\mu  )
\eeq
The formulae (2.13) for twisted coproduct is modified as follows:\\
\parbox{12cm}{
\beqq
\Delta^F (M_{\mu \nu  }) \rightarrow \Delta^F (M_{\mu \nu  })+\rho^\mu
_+ (P_\mu  \otimes Z + Z\otimes P_\mu  ) +
\eeqq
\beqq
+\rho^\mu _- (P_\mu \otimes Z - Z\otimes P_\mu )
\eeqq}\parbox{1.5cm}{\beq \eeq}\\
With the choice (2.16) the explicite formulae for the universal $R-$matrix is
the following:\\
\parbox{12cm}{
\beqq
R=\exp (-2f_-^{(z)})=\exp (-2\rho^{\mu \nu  }_- (P_\mu \otimes P_\nu   - P_\nu
\otimes P_\mu  ))=
\eeqq
\beqq
\exp( -2\rho^\mu _- (P_\mu \otimes Z - Z\otimes P_\mu  ))
\eeqq}\parbox{1.5cm}{\beq \eeq}\\
The invariant tensor (2.3) takes the form
\beq
U=\exp (-2\alpha_+^{\mu \nu  }P_\mu \cdot P_\nu   -2\rho^\mu _+ P_\mu \cdot Z)
\eeq
and using the formulae $S^F=US_0U^{-1}$ one gets\\
\parbox{12cm}{
\beqq
S^{F^{(z)}}(P_\mu )=S_0(P_\mu )=-P_\mu
\eeqq
\beqq
S^{F^{(z)}}(M_{\mu \nu  })=-M_{\mu \nu  }-2(\alpha_{+\mu }\;^\rho
P_\rho -\alpha_\nu  \;^\rho P_\mu P_\rho)
\eeqq
\beqq
-(\rho_{+\mu }P_\nu   -\rho_{+\nu}P_\mu )\cdot Z
\eeqq}\parbox{1.5cm}{\beq \eeq}\\
The reality conditions for the parameters $\rho_{\pm}^\mu $ take the form:\\
\parbox{12cm}{
a) + -- involution: $\rho^\mu _\pm$ real\\
b) $\oplus$ -- involution: $(\rho^\mu _+)^*=\rho^\mu _- $}\parbox{1.5cm}{\beq
\eeq}\\
In this Section we considered classical twisted \pr\ algebras,
parametrized by multiparameter twist functions. These Hopf algebras by
duality relations determine multiparameter deformations of the
functions of the \pr\ group. Using the duality relation between
multiplication and comultiplication
\beq
<a\cdot b,c> = <a\otimes b, \Delta(c)>
\eeq
one sees easily that all the antisymmetric contributions to the
twisted coproducts (see e.q.(2.13)) lead to noncommutativity of the
generators of the corresponding dual quantum \pr\ group.\\
It is an interesting exercise to classify the quantum \pr\ groups
dual to the classical twisted \pr\ algebras.\\

\addtocounter{chapter}{1}
\section{An example: Chaichian-Demiczev quan\-tum \pr\ algebra}
\setcounter{equation}{0}
We shall show that the example of $q-$\pr\ algebra given in [10] is
isomorphic as a Hopf algebra to twisted classical \pr\ algebra.
We shall describe firstly the
complexified classical Lorentz algebra $SO(4;C)=SO(3;C)\oplus SO(3;C)$ as
follows:\\
\parbox{12cm}{
\beqq
[e_i,e_{-j}]=\delta_{ij}h_i
\eeqq
\beqq
[h_i,h_j]=0
\eeqq
\beqq
[h_i,e_{\pm j}]=\pm 2\delta_{ij}e_{\pm j}
\eeqq}\parbox{1.5cm}{\beq \eeq}\\
where $(e_1,e_{-1},h_1)$ and $(e_2,e_{-2},h_2)$ describe two $O(3;C)$ sectors.

Introducing\\
\parbox{12cm}{
\beqq
L_1=e_{-1}\quad\quad\quad L_2=e_{-2}\quad\quad\quad L_5={1 \over
2}(h_1+h_2) \eeqq
\beqq
L_3=e_{+2}\quad\quad\quad  L_4=e_{+1}\quad\quad\quad L_6={1\over
2}(h_2-h_1) \eeqq}\parbox{1.5cm}{\beq \eeq}\\
one obtains the relations\\
\beqq
[L_1,L_5]=L_1 \quad\quad\quad\quad [L_2,L_5]=L_2
\eeqq
\beqq
[L_1,L_6]=-L_1\quad\quad\quad\quad [L_2,L_6]=L_2
\eeqq
\parbox{12cm}{
\beqq
[L_1,L_4]=L_6-L_5 \quad\quad\quad\quad [L_2,L_3]=L_6+L_5
\eeqq
\beqq
[L_3,L_5]=-L_3 \quad\quad\quad\quad [L_4,L_5]=-L_4
\eeqq
\beqq
[L_3,L_6]=-L_3 \quad\quad\quad\quad [L_4,L_6]=L_4
\eeqq
\beqq
[L_1,L_2]=[L_1,L_3]=[L_2,L_4]=[L_3,L_4]=[L_5,L_6]=0
\eeqq}\parbox{1.5cm}{\beq \eeq}\\
where of course $(a=1\ldots 6)$
\beq
\Delta(L_a)=L_a\otimes I + I\otimes L_a
\eeq
Let us perform the twist of this coproduct
\beqq
F=q^{h_2\otimes h_1} =q^{(L_5+L_6)\otimes (L_5-L_6)}
\eeqq
One gets $(\Delta^F(L_a)=F\cdot \Delta (L_a)\cdot F^{-1})$\\
\parbox{12cm}{
\beqq
\Delta^F(L_1)=L_1\otimes I + q^{-2(L_5+L_6)}\otimes L_1
\eeqq
\beqq
\Delta^F(L_2)=I\otimes L_2 + L_2\otimes q^{-2(L_5-L_6)}
\eeqq
\beqq
\Delta^F(L_3)=I\otimes L_3 + L_3\otimes q^{2(L_5-L_6)}
\eeqq
\beqq
\Delta^F(L_4)=L_4\otimes I + q^{2(L_5+L_6)}\otimes L_4
\eeqq
\beqq
\Delta^F(L_5)=\Delta (L_5)\quad\quad\quad\quad \Delta^F(L_6)=\Delta (L_6)
\eeqq}\parbox{1.5cm}{\beq \eeq}\\
\newpage
Introducing\\
\parbox{12cm}{
\beqq
\tilde{L}_1=L_1\quad\quad\quad
\tilde{L}_2=q^{-2}L_2q^{-2(L_5-L_6)}\quad\quad\quad
\tilde{L}_3=q^{-2}L_3q^{2(L_5-L_6)} \eeqq
\beqq
\tilde{L}_4=L_4\quad\quad\quad \tilde{L}_5=L_5\quad\quad\quad\tilde{L}_6=L_6
\eeqq}\parbox{1.5cm}{\beq \eeq}\\
one can identify the transformed classical Lorentz algebra (3.3) with
the $q-$deformed Lorentz
algebra proposed in [10], with the coproduct\\
\parbox{12cm}{
\beqq
\Delta^F(\tilde{L}_1)=\tilde{L}_1\otimes I +
q^{-2(\tilde{L}_5+L_6)}\otimes \tilde{L}_1
\eeqq
\beqq
\Delta^F(\tilde{L}_2)=I\otimes \tilde{L}_2 + \tilde{L}_2\otimes
q^{-2(\tilde{L}_5-\tilde{L}_6)}
\eeqq
\beqq
\Delta^F(\tilde{L}_3)=I\otimes \tilde{L}_3 + \tilde{L}_3\otimes
q^{2(\tilde{L}_5-\tilde{L}_6)}
\eeqq
\beqq
\Delta^F(\tilde{L}_4)=\tilde{L}_4\otimes I +
q^{2(\tilde{L}_5+\tilde{L}_6)}\otimes \tilde{L}_4
\eeqq
\beqq
\Delta^F(\tilde{L}_5)=\tilde{L}_5 \otimes I + I\otimes \tilde{L}_5
\eeqq
\beqq
\Delta^F(\tilde{L}_6)=\tilde{L}_6 \otimes I + I\otimes \tilde{L}_6
\eeqq}\parbox{1.5cm}{\beq \eeq}\\
Introducing fourmomentum operators, which in the basis (3.2) will satisfy the
following covariance
relations with $L_5$, $L_6$\\
\parbox{12cm}{
\beqq
[P_1,L_5]=P_1\quad [P_2,L_5]=0\quad [P_3,L_5]=-P_3\quad [P_4,L_5]=0
\eeqq
\beqq
[P_1,L_6]=0\quad [P_2,L_6]=P_2\quad [P_3,L_6]=0\quad [P_4,L_6]=-P_4
\eeqq}\parbox{1.5cm}{\beq \eeq}\\
one obtains after the nonlinear transformation\\
\parbox{12cm}{
\beqq
\tilde{P}_1=q^{L_5-L_6}P_1\quad\quad\quad\quad
\tilde{P}_2=q^{L_5-L_6}P_2 \eeqq
\beqq
\tilde{P}_3=q^{L_6-L_5}P_3\quad\quad\quad\quad
\tilde{P}_4=q^{L_6-L_5}P_4 \eeqq}\parbox{1.5cm}{\beq \eeq}\\
the relations\\
\parbox{12cm}{
\beqq
[\tilde{P}_1,\tilde{P}_2]_{q^2}=[\tilde{P}_4,\tilde{P}_1]_{q^2}=
[\tilde{P}_2,\tilde{P}_3]_{q^2}=[\tilde{P}_3,\tilde{P}_4]_{q^2}=0 \eeqq
\beqq
[\tilde{P}_1,\tilde{P}_3]=[\tilde{P}_2,\tilde{P}_4]=0
\eeqq}\parbox{1.5cm}{\beq \eeq}\\
and the coproducts\\
\parbox{12cm}{
\beqq
\Delta^F(\tilde{P}_1)=\tilde{P}_1\otimes 1
+q^{-2L_6}\otimes \tilde{P}_1
\eeqq
\beqq
\Delta^F(\tilde{P}_2)=\tilde{P}_2\otimes 1
+q^{2L_5}\otimes \tilde{P}_2
\eeqq
\beqq
\Delta^F(\tilde{P}_3)=\tilde{P}_3\otimes 1 +q^{2L_6}\otimes
\tilde{P}_3 \eeqq
\beqq
\Delta^F(\tilde{P}_4)=\tilde{P}_4\otimes 1 +q^{-2L_5}\otimes
\tilde{P}_4 \eeqq}\parbox{1.5cm}{\beq \eeq}\\
The relations (3.10-11) describe the translation sector of Chaichian--Demiczev
quantum algebra.

Let us recall that recently the quantum Lorentz
groups have been classified by Worononowicz
and Zakrzewski [11], where besides the Drinfeld--Jimbo
parameter $q$ a new parameter $t$ has been introduced. It can be shown that the
quantum deformation, proposed by Chaichian and Demiczev corresponds to $q=1$.
This
condition as the necessary requirement for the existence of nontrivial quantum
deformation
of \pr\ algebra, with the Lorentz part as the Hopf subalgebra, has been
obtained
in [12] (see also [13]).

It should be stressed that in [11] there were given also other examples of the
quantum deformations of the Lorentz group, which satisfy the condition $q=1$
and can be extended to the quantum deformations of the \pr\ algebra without
supplementing an
eleventh dilatation generator. It would be interesting to prove the conjecture
that all quantum deformations of
\pr\ algebra which do have the deformed Lorentz algebra as its Hopf subalgebra
are classical twisted \pr\ algebras.

We would like finally to mention that it is possible to obtain the \pr\ quantum
group as well as \pr\
quantum algebra with Drinfeld--Jimbo deformation parameter $q\neq 1$ if we
assume braided structure
of the tensor products, i.e. we consider the deformations in the framework of
braided quantum groups
and algebras (see e.g. [14]). In such a case the parameter $q$ enters into the
definition of braided tensor product of the Lorentz generators
and  the  translation generators [12] (see also ref. [15,16]). In this paper we
assume
however the standard ``bosonic'' relations for the tensor categories.
\addtocounter{chapter}{1}
\section{Discussion}
\setcounter{equation}{0}

i) Representation theory of twisted \pr\ algebra.

The theory of irreducible representations of  twisted \pr\ algebras is
described by the conventional Wigner representations for the \pr\ algebra
[1,2].
The twisting can be interpreted as the modification of the tensor products for
relativistic free particle states, in particular the 2-particle sectors in a
relativistic Fock space.
The tensor product $|1>\otimes |2>$ of two free one-particle states (i=1,2)
\beq
|i>=|m^{(i)},s^{(i)};p_\mu^{(i)},s_3^{(i)}>
\eeq
one modifies as follows
\setcounter{popnr}{\value{equation}}
\addtocounter{popnr}{1}
\alpheqn
\beq
|1>\otimes_F |2>=F(c^{(1)},c^{(2)})|1>\otimes |2>
\eeq
where $(\alpha=\alpha_++\alpha_-)$
\beq
F(c^{(1)},c^{(2)})=\exp \alpha_{ij}c^{(1)}_i c^{(2)}_j
\eeq
\reseteqn
Let us denote by $\hat{\alpha}$ the algebra describing the levels of the
representation
space (for (4.1) $\hat{\alpha}=(P_\mu,S_3),$ where
$S_\mu=1/2\epsilon_{\mu\nu\rho\tau}M^{\nu\rho}P^\tau)$,
and by $\hat{O}$ the Casimirs parametrizing by its eigenvalues the
representations
$(\hat{O}=(P_\mu P^\mu,S_\mu S^\mu)$ for ${\cal P}_4).$ One can
distinguish the following two cases:

i1) $[c_i,\hat{\alpha}]=0.$

This corresponds to our choice d) (see (2.11), (2.13)). In such a case the
twisted tensor
product of two representations (4.1) describe the fixed fourmomenta components
of the wave packet
\beq
|1,2>_F=\exp (\alpha^{\mu\nu}p_\mu^{(1)}p_\nu^{(2)})|1>\otimes|2>
\eeq
For dimensional reasons one should put
$\alpha^{\mu\nu}=\frac{1}{\kappa^2}a^{\mu\nu}$
($\kappa-$masslike parameter). If we assume that $a^{\mu\nu}$ has negative
eigenvalues, one
obtains from (4.3) the Gauss-like 2-particle wave function.\\

i2) $[c_i,\hat{\alpha}]\neq 0$\\

Such  a case is decribed by the choices a), b), c) of the twist function as
well
as
the example described in Sect.3. In such a case twisted two-particle states
described by (4.2)
are not eigenvalues of the ``two-particle observable''
$\Delta^F(\hat{\alpha}),$ because
\beq
\Delta^F(\hat{\alpha})=F\cdot \Delta(\hat{\alpha})\cdot F^{-1}\neq
\Delta (\hat{\alpha})
\eeq
For the fourmomentum operators the additivity of the fourmomenta eigenvalues is
modified by the formula
\beq
\Delta^F(P_\mu)=F\cdot (P_\mu\otimes 1 + 1\otimes P_\mu )\cdot F^{-1}
\eeq
In our example in Sect.3 the formulae (4.5) take the form (3.11).
The physical interpretation of generalized wave packets (4.2a) with modified
addition for the fourmomenta is not clear.

ii) Twisted \pr\ algebra from the contraction of $U_q(O(4,2)).$

In recent paper [17] two of the present authors proposed the contraction of
$U_q(O(4,2))$ to
quantum \pr\ algebra.  It can be shown
that the result of the contraction is a twisted \pr\ algebra with the twist
function
depending on the fourmomenta and one central charge $Z$ (see (2.16)), obtained
from the contraction
of the dilatation generator in the conformal algebra.\\

iii) Nonabelian choice of twist functions.

It is interesting to consider more general classes of twisting functions,
with $F$ spanned by nonabelian sectors of the algebra. In particular such a
twisting
function is provided by the universal $R-$matrix, which interchanges two
noncocomutative
coproducts $\Delta$ and $\Delta ' =\sigma\cdot\Delta$ of a quantum algebra.
It is known that for Drinfeld--Jimbo deformations $U_q(\hat{g})$ of
simple Lie algebras the universal
$R-$matrix can be decomposed into the product [18,19]
\beq
R=\prod_{\alpha\in\Delta^{(+)}} R_\alpha \cdot K
\eeq
where
\beq
R_\alpha =\exp q_\alpha (a_\alpha (q) e_\alpha \otimes e_{-\alpha})
\eeq
and $K$ depends only on the Cartan generators. It appears that any component
(4.7) of the product (4.6)
can be used as a twist function F [20]. Because $a_\alpha (q)$ is proportional
to $q-q^{-1},$ the twisting with $F=R_\alpha$ can be introduced only for
genuine
quantum algebras $(q\neq 1)$. It is interesting to find nontrivial twist
functions for quantum
$\kappa-$\pr\ algebra, proposed in [21,22]. Because the universal
$\hat{R}-$matrix
for $\kappa -$ \pr\ algebra is not known, the type of twisting proposed in
[20] can not be applied.
\\[1cm]
\newpage
{\bf Acknowledgements}\\
Two of the authors (J.L. and V.N.T.) would like to thank the University of
Geneve
for its generous hospitality. Third author (A.N.) would like to thank the
University of Bonn
and Prof. R. Flume for the warm hospitality, and DAAD for the financial
support.\\[2cm]
{\bf References}
\begin{enumerate}
\item E.P. Wigner, Ann. Math.{\bf 40}, 149 (1939)
\item A.S. Wightman, 1960 Les Houches Summer School, "Relations de
dispersion et particules elementaries", p.161
\item V.G. Drinfeld, Quantum groups, Proc.Intern.Congress of Mathematics,
(Berkeley,USA, 1986) p.798
\item L. Faddeev, N. Reshetikhin and L. Takhtajan, Alg.Anal.{\bf 1}, 178
(1989)
\item V.G. Drinfeld, Leningrad Math. Journ. 1, 1419 (1990)
\item N. Reshetkhin, Lett. Math. Phys. {\bf 20}, 331 (1990)
\item G. Gurevich and S. Maijd, ``Branded Groups of Hopf Algebras obtained by
Twistig'',
Cambridge Univ. preprint DAMTP 91-49
\item B. Enriquez, Lett. Math. Phys. {\bf 25}, 111 (1992)
\item A. Kempf, ``Multiparameter R-matrices, Subquantum Groups and
Ge\-ne\-ra\-li\-zed Twisting Method'',
M\"{u}nchen Univ. prep. LMU-TPW 91-4
\item M. Chaichian and A.P. Demiczev, ``Quantum \pr\ Group, Algebra and Quantum
Geometry
of Minkowski Space'', Helsinki Univ. preprint HU-TFT-93-24, March 1993
and Phys.Lett., {\bf B304}, 220 (1993)
\item S.L. Woronowicz and S. Zakrzewski, ``Quantum Deformations of Lorentz
group: Hopf
$*$-algebra level'', Warsaw Univ. preprint 1992, Compositio Mathematica,
in press
\item S. Majid, Journ. Math. Phys. {\bf 34}, 2045 (1993)
\item P. Podle\'{s}, and S.L. Woronowicz, private communication
\item S. Majid, J. Math. Phys. {\bf 34}, 1176 (1993)
\item J. Rembieli\'{n}ski,"Quantum braided \pr\ group", Lodz Univ.
preprint KFT UL 7/93
\item J.A. de Azcaraga, P. Kulish and F. Rodenas,"Reflection equations
and q-Minkowski space algebras", hep-th 9309036
\item J. Lukierski, A. Nowicki, Phys. Lett. {\bf B279}, 299 (1992)
\item S.M. Khoroshkin and V.N. Tolstoy, Comm. Math. Phys. {\bf 141},
559 (1991)
\item S. Levendorskii, Y.Soibelman, Comm.Math.Phys., {\bf 139}, 141
(1991)
\item S.M. Khoroshkin and V.N. Tolstoy,"Twisting of quantum
(super)algebras. Connection of Drinfeld's and Cartan-Weyl realizations
for quantum affine algebras", preprint MPI, Bonn 1993
\item J. Lukierski, A. Nowicki, H. Ruegg and V.N. Tolstoy, Phys. Lett. {\bf
B264},
331 (1991)
\item J. Lukierski, A. Nowicki and H. Ruegg, Phys.Lett. {\bf B293}, 344 (1992)
\end{enumerate}
\end{document}